# ON EQUIVALENCE OF THINNING FLUIDS USED FOR HYDRAULIC FRACTURING

Alexander M. Linkov

*Institute for Problems of Mechanical Engineering, 61, Bol'shoy pr. V. O., Saint Petersburg, 199178, Russia*
*Presently: Rzeszow University of Technology, ul. Powstancow Warszawy 8, Rzeszow, 35-959, Poland,*
*e-mail: linkoval@prz.edu.pl*

*Abstract.* The paper aims to answer the question: if and how non-Newtonian fluids may be compared in their mechanical action when used for hydraulic fracturing? By employing the modified formulation of the PKN problem we obtain its simple analytical solutions in the cases of perfectly plastic and Newtonian fluids. Since the results for shear thinning fluids are intermediate between those for these cases, the obtained equation for the fracture length suggests a criterion of the equivalence of various shear thinning fluids for the problem of hydraulic fractures. We assume fluids equivalent in their hydrofracturing action, when at a reference time they produce fractures of the same length. The equation for the fracture length translates the equivalence in terms of the hydraulic fracture length and treatment time into the equivalence in terms of the properties of a fracturing fluid (behavior and consistency indices). Analysis shows that the influence of the consistency and behavior indices on the fracture length, particle velocity and propagation speed is quite similar when considering the PKN and KGD models despite the models refer to different plain-strain states. It appears that from the mechanical point of view, the differences between thinning fluids are not significant; they may be taken into account when wishing to have some quantity to be greater (less) at small (large) time. We conclude that a choice of a fracturing fluid is to be made mostly from technological and/or economic considerations under the condition that the compared fluids are equivalent in their mechanical effect according to the suggested criterion.

## 1. INTRODUCTION

Hydraulic fracturing is widely used for increasing production of oil, gas and thermal reservoirs. Since the pioneering works by Khristianovich & Zheltov [1], Perkins & Kern [2], Geertsma & de Klerk [3] and Nordgren [4], it has been a subject of numerous investigations (see, e. g. reviews in the papers [5-9]). Theoretical investigations concerned mostly with asymptotics near the fluid front and regimes of the fracture propagation (e.g., [10-14]). Benchmark solutions of particular problems have been given in [4, 13, 15] for the PKN model and in [16] for the KGD model when the fracturing fluid is *Newtonian*. In view of practical significance of non-Newtonian fluids (see, e. g. [17, 18]), the influence of non-linear dependence between the shear strength $\tau$ and the shear strain rate $\dot{\gamma}$ was studied by Adachi & Detournay [5]. The authors considered the power reological law

$$\tau = M\gamma^n, \qquad (1)$$

where $M$ is the consistency index and $n$ is the fluid behavior index. They solved numerically the boundary value problem (BVP) for the geometrical scheme of KGD model under the assumption of zero fracture resistance ($K_{IC} = 0$) and zero lag. To compare various fluids, the authors used the apparent (reference) dynamic viscosity $\mu_r$ of an 'equivalent' Newtonian fluid ($n_r = 1$) and prescribed an arbitrary (reference) value $\dot{\gamma}_r$ of the shear strain rate. Then the equality of shear strengths, expressed by $M\dot{\gamma}_r^{\,n} = \mu_r \dot{\gamma}_r$, yields the consistency index $M$ for the 'equivalent' (under the selected $\mu_r$ and $\dot{\gamma}_r$) non-Newtonian fluid:

$$M = \mu_r \dot{\gamma}_r^{1-n}. \tag{2}$$

By setting $\mu_r = 0.1$ Pa s, $\dot{\gamma}_r = 50$ 1/s, it was established that fractures driven by fluids with lower index $n$ propagated faster at early time and slower at large time, than the fractures driven by fluids with higher index $n$. The authors conclude [5, p. 596]: "This apparent paradox is explained by the fact that the viscosity of a power-law fluid depends not only on its values $M$ and $n$, but more importantly on the relative position of its current state along the reology curve." This profound conclusion raises the question: if and how it is possible to choose the reference shear strain $\dot{\gamma}_r$ to compare non-Newtonian fluids in their mechanical action as regards hydraulic fracturing? The present paper aims to answer the question.

The objective is reached on the basis of the modified formulation of the hydraulic fracture problem [9], which provides simple analytical solutions to the PKN and KGD models. The modified formulation is used (Sec. 2) to extend the solution of the PKN problem to a non-Newtonian fluid, described by the power-law (1). Starting from Sec. 3 we focus on shear thinning ($0 < n < 1$) fluids, commonly used for hydraulic fractures. We employ the fact that the results for $0 < n < 1$ are intermediate between those for perfectly plastic ($n = 0$) and Newtonian ($n = 1$) fluids. It appears that the solution for a shear thinning fluid in *self-similar* variables is practically independent on its properties. This gives simple universal formulae for the self-similar fracture length, fracture speed, particle velocity, opening, pressure and flux. In Sec. 4, the formulae are used to analyze the evolution of the *dimensional* quantities in time. This provides us with a key for comparing shear thinning fluids in their mechanical action as regards hydraulic fractures. We define fluids 'equivalent' in their action, when at a reference time of hydraulic treatment they produce fractures of the same length. The derived equation for the fracture length serves us to translate this 'equivalence in hydrofracturing effect' into the 'equivalence in fluid properties'. Actually, this gives us the needed reference rate $\dot{\gamma}_r$ to use equation (2). It appears that although the results are obtained by using the PKN model, they are applicable to the KGD model, as well. Sec. 5 summarizes the conclusions obtained.

## 2. MODIFIED FORMULATION AND SELF-SIMILAR SOLUTION OF PKN PROBLEM

*2.1. Modified formulation of PKN problem*

We consider the geometrical scheme of the PKN model (Fig. 1), for which plane-strain conditions occur in cross sections parallel to the fracture front. Then the elasticity equations yield the proportionality of the fracture opening $w$ to the net pressure $p$ [4]:

$$p = k_s w, \tag{3}$$

where $k_s = (2/\pi h) E /(1-\nu^2)$, $E$ is the elasticity modulus, $\nu$ is the Poisson's ratio, $h$ is the fracture height. The hydraulic fracture is driven by a fluid with the power-law (1). Then for flow in a narrow ($w \ll h$) channel, usual derivation of the Poiseuille equation gives the fluid particle velocity averaged across the channel opening:

$$v = \left(-k_f w^{n+1} \frac{\partial p}{\partial x}\right)^{1/n}, \tag{4}$$

where $k_f = 1/(\theta M)$; for an elliptic channel with the axes $w$ and $h$, we have $\theta = [\pi(1+\pi n - n)/(2n)]^n$ (e.g. [19]); for a thin plane channel, $\theta = 2[2(2n+1)/n]^n$ (e.g. [5]). The ratio of these values is close to the unity being 1.0 for a perfectly plastic fluid ($n = 0$) and $12/\pi^2$ for a Newtonian fluid ($n = 1$). For certainty, we shall use the value for a plane channel; then $\theta = 2$ for a perfectly plastic fluid with the shear strength $M = \tau_0$ and $\theta =$ 12 for a Newtonian fluid with the dynamic viscosity $M = \mu$. Note that by definition the flux $q$ per the channel opening is $q = vw$.

We use the modified formulation of the lubrication equation in terms of the particle velocity $v$ and the modified opening $y = w^{1/\alpha}$, where α is defined by the condition that the particle velocity is neither non-zero nor infinite at the fluid front. For the considered PKN model, substitution (3) into (4) and taking into account that the opening is zero at the front imply that the function $y = w^{n+2}$ should be linear near the front. Hence, $\alpha = 1/(n+2)$.

For simplicity we neglect leak-off into formation. Then the modified lubrication equation [9] reads:

$$\frac{\partial v}{\partial x} + \frac{\alpha}{y}\frac{\partial y}{\partial x}v + \frac{\alpha}{y}\frac{\partial y}{\partial t} = 0, \tag{5}$$

where, as follows from (3) and (4), the dependence between $y$ and $v$ is:

$$v = \left(-k_f k_s \alpha \frac{\partial y}{\partial x}\right)^{1/n}. \tag{6}$$

We seek the solution of the PDE (5), where the dependence between $y$ and $v$ is given by (6), under the initial condition of zero opening along a perspective fracture path

$$y(x,0) = 0, \tag{7}$$

the boundary condition (BC) of the prescribed influx $q_0(t)$ (per unit height) at the inlet ($x = 0$):

$$y^\alpha v\big|_{x=0} = q_0(t). \tag{8}$$

and the BC of zero opening at the fracture front $x = x_*$:

$$y(x_*,t) = 0. \tag{9}$$

Besides, the solution satisfies the speed equation (SE) [20], which expresses that the particle velocity at the front $v(x_*,t)$ equals the fracture propagation speed $v_*(t)$:

$$v(x_*,t) = v_*(t) = \left(-k_f k_s \alpha \frac{\partial y}{\partial x}\right)^{1/n}\bigg|_{x=x_*}. \tag{10}$$

As shown in [21], if a solution of (5) with $v$ defined by (6) meets the BC (9), it identically satisfies the SE (10). Hence, for a fixed position of the fracture front, we actually have two BC (9) and (10) at the front rather than one condition (9). This makes the BV problem (5)-(9) *ill-posed* for any fixed position of the front $x_*$ [20, 21]. Below to avoid complications, we shall solve the initial value problem (5), (6), (9), (10), instead of the BVP (5), (6), (8), (9) for a fixed front position of the front $x_*$. Substitution of the solution into the BC at the inlet (8) gives the corresponding influx $q_0$. By changing $x_*$, we may find that value of $x_*$, for which the BC (8) is met to a prescribed tolerance.

*2.2. Normalized variables. Formulation not including consistency index*

We exclude the consistency index $M$ from the problem formulation by using the dimensionless (normalized) variables:

$x_d = x/x_n$, $x_{*d} = x_*/x_n$, $t_d = t/t_n$, $v_d = v/v_n$, $v_{*d} = v_*/v_n$, $w_d = w/w_n$, $p_d = p/p_n$,

$$q_d = q/q_n, \ q_{0d} = q_0/q_n, \ y_d = y/y_n, \tag{11}$$

where the normalizing quantities are chosen as

$$x_n = (k_f k_s q_n^{n+2} t_n^{2n+2})^{\frac{1}{2n+3}}, \ w_n = q_n t_n/x_n, \ p_n = k_{rn}w_n, \ v_n = x_n/t_n, \ y_n = w_n^{1/\alpha} \tag{12}$$

with $t_n$ and $q_n$ being typical values of the time and influx per unit height, respectively. In terms of the normalized variables the problem formulation (5)-(10) has the same form with the only difference: the multiplier $k_f k_s$ is changed to the unity what excludes the consistency factor. Henceforth, when there may be no confusion, we shall use (5)-(10) for the normalized variables omitting the subscript '$d$' in their notation.

*2.3. Formulation in self-similar variables. Analytical solution for non-Newtonian fluids*

Consider the case when the influx at the inlet is prescribed by the power dependence in time. In terms of the normalized variables the influx is:

$$q_0(t) = t^{\beta_q}, \qquad (13)$$

where $\beta_q$ is a dimensionless constant; note that $\beta_q = 0$ for constant influx. The solution of (5)-(10) may be found by using the self-similar variables defined by equations:

$$x = \xi t^{\beta_*}, \quad x_* = \xi_* t^{\beta_*}, \quad v = V(\xi) t^{\beta_* - 1}, \quad v_* = V_* t^{\beta_* - 1}, \qquad (14)$$

$$w = W(\xi) t^{\beta_w}, \quad y = Y(\xi) t^{\beta_w / \alpha}, \quad p = P(\xi) t^{\beta_p}, \quad q = Y(\xi)^\alpha V(\xi) t^{\beta_q},$$

where $\xi_*$ and $V_* = \xi_* \beta_*$ are constants, expressing the self-similar fracture length and the self-similar fracture speed, respectively. As $x / x_* = \xi / \xi_*$, the self-similar coordinate $\xi = \xi_* x / x_*$ actually gives the distance from the inlet normalized by the fracture length $x_*$. Thus, in fact, the formulae (14) represent the solution in the form of products of functions with separated variables $\varsigma = \xi / \xi_* = x / x_*$ and $t$. Substitution of (14) into (5)-(10) yields that the powers of time cancel when

$$\beta_w = \beta_p = \frac{1 + (n+1)\beta_q}{2n+3}, \quad \beta_* = \frac{2(n+1) + (n+2)\beta_q}{2n+3}. \qquad (15)$$

Then the PDE (5) turns into the ordinary differential equation in the self-similar variables:

$$\frac{dV}{d\xi} + \alpha \frac{V(\xi) - V_* \xi / \xi_*}{Y(\xi)} \frac{dY}{d\xi} + \beta_w = 0 \qquad (16)$$

with the dependence between $V$ and $Y$, following from (6),

$$V(\xi) = \left(-\alpha \frac{dY}{d\xi}\right)^{1/n}. \qquad (17)$$

The initial condition (7) is met identically if $\beta_q > -1/(n+1)$. Further on, we assume this inequality fulfilled, focusing on the case of constant influx when ($\beta_q = 0$).

The BC (8) and (9) become, respectively,

$$Y^\alpha V(0) = A, \qquad (18)$$

$$Y(\xi_*) = 0. \qquad (19)$$

Actually, in (18), $A = 1$. We have written $A$ for further discussion of the solution.

The SE (10) reads

$$V_* = \xi_* \beta_* = \left(-\alpha \frac{dY}{d\xi}\right)^{1/n} \bigg|_{\xi = \xi_*}. \qquad (20)$$

For any fixed $\xi_*$, the analytical solution of the *initial value* (Cauchy) problem (16), (17), (19), (20) is found in rapidly converging series (see Appendix). When having $Y$ and $V$, the BC (18) defines the self-similar influx $A$, corresponding to the accepted $\xi_*$. By changing $\xi_*$, we may find that value of $\xi_*$, for which the BC (18) is met to a prescribed tolerance when $A = 1$.

Actually, there is no need in solving the problem for various $\xi_*$. It can be checked by direct substitution that if $Y_1(\xi)$ is the solution for $\xi_* = \xi_{*1}$ so that the corresponding influx is $A = A_1$, then the solution for an arbitrary influx $A$ is given by equations:

$$\xi_* = \xi_{*1}\left(\frac{A}{A_1}\right)^{(n+2)/(n+3)}, \qquad Y(\xi) = \left(\frac{\xi_*}{\xi_{*1}}\right)^{n+1} Y\left(\xi \frac{\xi_{*1}}{\xi_*}\right). \tag{21}$$

In particular, when $\xi_{*1}=1$, $A = 1$, equations (21) become: $\xi_* = A_1^{-(n+2)/(n+3)}$, $Y(\xi) = \xi_*^{n+1} Y(\xi/\xi_*)$. Hence, it is sufficient to find the solution for $\xi_{*1} = 1$.

## 3. UNIFIED SELF-SIMILAR SOLUTION FOR THINNING FLUIDS

Shear thinning fluids have the behavior index ($0 < n < 1$) intermediate between those for the limiting cases of perfectly plastic ($n = 0$) and Newtonian ($n = 1$) ones. Therefore, by continuity, we may infer conclusions for thinning fluid from the results of the limiting cases. It appears that the solutions in self-similar variables for $n = 0$ and $n = 1$ are quite close, so that all the thinning fluids may be described by universal equations, not depending on the behavior index.

Consider, for certainty, the case of constant influx ($\beta_q = 0$).

Then *for a perfectly plastic fluid* ($n = 0$) the solution is (see equations (A4) in Appendix): $\xi_{*P} = (9/8)^{1/3} = 1.04004$, $V_{*P} = (2/3)\xi_{*P} = 0.693361$, $Y_P(\xi) = 2(\xi_{*P} - \xi)$, $V_P(\xi) = V_{*P}$.

*For a Newtonian fluid* ($n = 1$), the solution is given in [9] (see also Appendix for this particular case). It is: $\xi_{*N} = 1.00101$, $V_{*N} = (4/5)\xi_{*N} = 0.800808$, $Y_N(\xi) = (12/5)(\xi_{*N})^2 f_Y(\xi/\xi_{*N})$, $V_N(\xi) = V_{*N} f_V(\xi/\xi_{*N})$, where functions $f_Y(\varsigma)$ and $f_V(\varsigma)$ are defined by rapidly converging series $f_Y(\varsigma) = \sum_{j=1}^{\infty} a_j (1-\varsigma)^j$, $f_V(\varsigma) = \sum_{j=0}^{\infty} b_j (1-\varsigma)^j$, with $a_1 = b_0 = 1$, $b_1 = -1/16$ and other coefficients ($j \geq 2$) evaluated recurrently by using equations:

$$a_j = \frac{b_{j-1}}{j}, \quad b_j = -\frac{1}{j+1/3}\left[\frac{4j-3}{12}a_j + \sum_{k=2}^{\infty}\left(j+1-\frac{2}{3}k\right)a_k b_{j-k+1}\right].$$

At the inlet ($\xi = 0$), this gives: $f_Y(0) = 0.970106$, $f_V(0) = 0.941528$, $Y(0) = 2.332959$, $V(0) = 0.753983$. Therefore, to the accuracy of three percent, the self-similar particle velocity is almost constant with the average value $V_N(\xi) = 0.78$, while the self-similar modified opening is almost linear $Y_N(\xi) = 2.33(\xi_{*N} - \xi)$ along the fracture.

By comparing $\xi_{*N}$, $V_N$ and $Y_N(\xi)$ with analogous values $\xi_{*P}$, $V_{*P}$ and $Y_P(\xi)$ for a perfectly plastic fluid we see that they do not differ significantly. By continuity, this implies that for shear thinning fluids, one may use the approximate self-similar solution not depending on the behavior index $n$:

$$\xi_* = 1.02, \ V = 0.74, \ Y(\xi) = 2.20(\xi_* - \xi). \tag{22}$$

The relative error of equations (22) does not exceed 2 percent in $\xi_*$, 7.6 percent in $V$ and 6 percent in $Y(\xi)$. Note that similar conclusion on small dependence of the self-similar solution on the behavior index follows also from the numerical results obtained in [5] for the KGD model despite significant difference in the asymptotic behavior of the solution near the fracture front as compared with the PKN model.

# 4. NORMALIZED AND DIMENSIONAL SOLUTIONS.
# COMPARISON OF NON-NEWTONIAN SHEAR THINNING FLUIDS

*4.1. Solution in normalized variables*

Using the solution (22) in the definitions (14) yields the normalized (dimensionless) values (now we write the subscript '$d$'):

$$x_{d_*}(t_d) = 1.02 t_d^{\beta_*}, \quad v_d(x,t) = v_{d_*}(t) = 1.02 t_d^{\beta_* - 1}, \quad y_d(x_d, t_d) = 2.24\left(1 - \frac{x_d}{x_{d_*}}\right) t_d^{\beta_w/\alpha}, \quad (23)$$

$$w_d(x_d, t_d) = p_d(x_d, t_d) = \left[2.24\left(1 - \frac{x_d}{x_{d_*}}\right)\right]^\alpha t_d^{\beta_w}, \quad q_d(x_d) = 1.02\left[2.24\left(1 - \frac{x_d}{x_{d_*}}\right)\right]^\alpha,$$

where in the considered case of constant influx, $\beta_* = 2(n+1)/(2n+3)$, $\beta_w = 1/(2n+3)$; as above, $\alpha = 1/(n+2)$. Recall that the normalized values do not depend on the consistency index. Now we see that the normalized fracture length, particle velocity, speed of propagation, opening, pressure and flux behave similarly for any *behavior index*. The difference is actually only in the exponents in time depending factors. The time exponents for a perfectly plastic fluids are $\beta_*$=2/3, $\beta_w$=1/3, $\alpha$=1/2; for a Newtonian fluid $\beta_*$=4/5, $\beta_w$=1/5, $\alpha$=1/3. Therefore, the difference in corresponding exponents for thinning fluids does not exceed 2/15 both for $\beta_*$ and $\beta_w$.

*4.2. Solution in physical variables. Criterion of equivalence*

The similarity of the solutions in the normalized variables, evident from (23), does not mean that non-normalized physical quantities also behave similarly. It is even impossible to compare non-Newtonian fluids if not making additional assumptions. Indeed, from the second of (23), it follows that the particle velocity is very large for small time, tending to infinity when $t \to 0$, and it is very small for large time, tending to zero when $t \to \infty$. It is easy to show that the shear strain rate $\dot{\gamma}$ is proportional (with the factor $\theta^{1/n}/w$) to the particle velocity. As a result, the shear strain rate $\dot{\gamma}$ takes values in the entire interval [0, ∞] what makes impossible to compare fluids by using equation (2) for the 'equivalent' consistency index.

Still by excluding too small and too large time, the comparison is possible. To obtain the needed rule for the choice of the reference value $\dot{\gamma}_r$, we consider the dimensional (physical) values. For them, from the definitions (11) and (14) it follows:

$$x_*(t) = \xi_*(k_f k_s q_0^{n+2})^{\beta_w} t^{\beta_*}, \qquad v_*(t) = \beta_* x_*(t) t^{-1}, \quad (24)$$

$$w(x,t) \approx \left[2.24\left(1 - \frac{x}{x_*}\right)\right]^\alpha \left(\frac{q_0^{n+1} t}{k_f k_s}\right)^{\beta_w}, \quad p(x,t) = k_s w(x,t), \quad q(x,t) \approx v_*(t) w(x,t),$$

where we have taken the constant value $q_0$ of the influx as the normalizing flux, and accounted for the second and forth of equations (23). The consistency index $M$ enters the solution (24) via the coefficient $k_f = 1/(\theta M)$.

The first of equations (24) suggests the needed *criterion* of fluid equivalence as concerned with hydraulic fracturing. We assume *fluids equivalent in their action in hydraulic fracturing when at a prescribed reference time $t_r$ (say, treatment time) they produce fractures of the same length.*

As mentioned, for thinning fluids, it is sufficient to consider the limiting cases of perfectly plastic and Newtonian fluids. By using the first of (24) for these cases and equating the results we obtain the reference shear strength $\tau_r = \tau_P$:

$$\tau_r = \left(\frac{\xi_{*P}}{\xi_{*N}}\right)^3 \left(54\frac{q_0 k_s^2 \mu_r^3}{t_r^2}\right)^{1/5}. \quad (25)$$

Under prescribed $k_s$, $q_0$ and $\mu_r$, equation (25) establishes the correspondence between the reference treatment time $t_r$ and the reference shear strength of a thinning fluid. Since $\tau_r = \mu_r \dot\gamma_r$, it can be also written as:

$$\dot\gamma_r = 54^{1/5}\left(\frac{\xi_{*P}}{\xi_{*N}}\right)^3 \left(\frac{k_s\sqrt{q_0}}{\mu_r t_r}\right)^{2/5}. \quad (26)$$

Equation (26) presents one-to-one correspondence between the reference treatment time $t_r$ and the reference shear strain rate $\dot\gamma_r$.

Above we have obtained $\xi_{*P} = 1.04004$, $\xi_{*N} = 1.00101$. By definition, $k_s = (2/\pi h)E/(1-\nu^2)$. When using the reference values of the influx $q_r$, fracture height $h_r$, elasticity modulus $E_r$ and Poisson's ratio $\nu_r$, equation (26) becomes:

$$\dot\gamma_r = 2.0790\left(\frac{E_r\sqrt{q_r}}{(1-\nu_r^2)h_r \mu_r t_r}\right)^{2/5}. \quad (27)$$

An alternative form of (27) employs a typical total influx $Q_r = q_r h_r$ instead of $h_r$:

$$\dot\gamma_r = 2.0790\left(\frac{E_r q_r^{3/2}}{(1-\nu_r^2)\mu_r Q_r t_r}\right)^{2/5}. \quad (28)$$

Equations (25)-(28) translate the equivalence of thinning fluids in terms of their action in hydraulic fracturing into the equivalence in terms of the fluid consistency index, defined by equation (2).

*4.3. Implications from the criterion. PKN versus KGN model*

The discussion above referred to the PKN model. To see if the conclusions following from equations (25)-(28) are general enough, it is reasonable to examine them against those, which may be deduced from the numerical results, obtained in the papers [5, 6] for the KGD model. Adachi & Detournay [5], assumed that the reference shear rate is $\dot\gamma_r$ =50 1/s and the reference Newtonian viscosity is that of water: $\mu_r = 0.1$ Pa·s. Thus the corresponding reference shear strength is $\tau_r = \mu_r \dot\gamma_r = 5$ Pa. These authors used the elasticity modulus $E_r$ = 2.5·10$^4$ MPa, Poisson's ratio $\nu_r = 0.15$, and the influx per unit height $q_r = 10^{-3}$ m$^2$/s. Following [5], Garagash [6] used the same reference values of $\dot\gamma_r$, $\mu_r$, $E_r$ and $\nu_r$, while the value of $q_r$ was taken two-fold less ($q_r$ = 0.5·10$^{-3}$ m$^2$/s). Below, to compare our results with those of [5, 6], we shall employ the same reference values $\mu_r = 0.1$ Pa·s, $E_r = 2.5\cdot10^4$ MPa, $\nu_r = 0.15$ and two values of the reference influx $q_r = 10^{-3}$ m$^2$/s and $q_r = 0.5\cdot10^{-3}$ m$^2$/s.

Firstly, we want to check, if the one-to-one correspondence between the reference time $t_r$ and the reference shear rate $\dot\gamma_r$ (or, equivalently, the reference shear strength $\tau_r$), established by (26) (or (25)), holds for the KGD model, as well. From the common derivations for the KGD model (see, e.g. [16, 5]), it is evident that, similar to the PKN model, introduction of properly normalized dimensionless variables excludes the consistency index, while using proper self-similar variables actually represents the solution in the form with separated variables. This results in equation for the fracture length, which has the form of equation (24) differing from it only by exponents and by the definition of the coefficient $k_s$. (The latter now does not contain the fracture height, which is not present in the KGD model). Then by employing the criterion of equal fracture length at a prescribed reference time and by using perfectly plastic and Newtonian fluids as the limiting cases for shear thinning fluids, one arrives at an equation of (26) type. (In the paper [5], it is equation (14)). Therefore, similar to the PKN model, there is one-to-one correspondence between the reference treatment time $t_r$ and the reference shear strain rate $\dot{\gamma}_r$.

This conclusion may be verified by an analysis of numerical results given in the papers [5, 6]. Note that the starting definition of the equivalence as the equality of the fracture lengths for perfectly plastic and Newtonian fluids at a reference time, by continuity, implies that the curves of the fracture length $x_*$ against time $t$ should intersect at the same point for any thinning fluid. The numerical results of the papers [5, 6] comply with this implication. Fig. 7 of the paper [5] shows that the fracture length $x_{*r}$, calculated for fluids with various behavior index $n$, was the same ($x_{*r} = 700$ m) at the time $t_r = 10^4$ s. Thus, actually, the results of the paper [5] correspond to the reference time $t_r = 10^4$ s. Below we shall use this value for quantitative comparison. Similar conclusion follows from Fig. 7a of the paper [6]: the fracture length $x_{*r}$, calculated for fluids with various behavior index $n$, was the same ($x_{*r} = 450$ m) at the time $t_r = 0.8 \cdot 10^4$ s. This implies that the results for the KGD model may be interpreted in terms of our definition of the equivalence.

Secondly, it is of interest to compare numerical values of the reference shear strain rate, following from (26), with that accepted in [5, 6] ($\dot{\gamma}_r = 50$ 1/s). For the values $\mu_r = 10^{-7}$ MPa·s, $E_r = 2.5 \cdot 10^4$ MPa, $\nu_r = 0.15$, used in [5, 6], equation (28) yields $\dot{\gamma}_r = 7.60298 \cdot 10^4 q_r^{3/5}/(Q_r t_r)^{2/5}$. With a typical total constant influx $Q_r = 0.05$ m$^3$/s and the reference time $t_r = 10^4$ s, we obtain $\dot{\gamma}_r = 100.3$ 1/s for $q_r = 10^{-3}$ m$^2$/s, accepted in [5], and $\dot{\gamma}_r = 66.2$ 1/s for $q_r = 0.5 \cdot 10^{-3}$ m$^2$/s, used in the paper [6]. The corresponding reference shear strength is $\tau_r = \mu_r \dot{\gamma}_r = 10.03$ Pa for $q_r = 10^{-3}$ m$^2$/s and $\tau_r = 6.62$ Pa for $q_r = 0.5 \cdot 10^{-3}$ m$^2$/s against the value 5 Pa, actually used in the papers [5, 6]. We see that for the same reference time $t_r = 10^4$ s, the *orders* of the reference shear strain rates and the shear strengths for the PKN-model and KGD-model are the same although there are differences in particular values.

The corresponding reference fracture length is $x_{*r} = 1215$ m for $q_r = 10^{-3}$ m$^2$/s and $x_{*r} = 698$ m for $q_r = 0.5 \cdot 10^{-3}$ m$^2$/s. These values agree with the Nordgren's results for a Newtonian fluid at the time $t_r = 10^4$ s (see eqn (C-9) of the paper [4]).

Thirdly, we may compare differences in evolution of hydrofracture quantities caused by the difference in the consistency and behavior indices. Write the dimensional fracture length, defined by (24) for an arbitrary thinning fluid, in terms of the reference values $x_{*r}$ and $t_r$ as $x_* = x_{*r}(t/t_r)^{\beta_*}$. Then the ratio of the fracture length for a Newtonian fluid $x_{*N}$ to that for a perfectly plastic fluid $x_{*p}$ is:

$$\frac{x_{*N}}{x_{*p}} = \left(\frac{t}{t_r}\right)^{\beta_d}, \qquad (29)$$

where $\beta_d = \beta_{*N} - \beta_{*p}$. For a constant influx, we have $\beta_{*N} = 4/5$, $\beta_{*p} = 2/3$, then the exponent in (29) is $\beta_d = 2/15$. Hence, the ratio $x_{*N}/x_{*p}$, being the unit at $t = t_r$, is less (greater) than the unit when $t < t_r$ ($t > t_r$). The ratio is very small, when $t \ll t_r$, and it is very large, when $t \gg t_r$. Still, as the exponent $\beta_d$ is quite small, the difference is not really great for the time within the range of practical significance (10 s $< t <$ 10$^5$ s). Specifically, for the accepted reference time $t_r = 10^4$ s, the ratio $x_{*N}/x_{*p}$ equals 0.398, 0.541, 0.736, 1.0 and 1.359 for the time instances 10 s, 100 s, 1000 s, $10^4$ s and $10^5$ s, respectively. The ratios $x_{*N}/x_{*p}$ for $t = 10$ s, 100 s, 1000 s and $10^4$ s may be compared with those, given in the papers [5, 6]. From log-log graphs in Fig. 7 of the paper [5], it may be inferred that approximate values of $x_{*N}/x_{*p}$ for these instances are 0.40, 0.54, 0.75 and 1.0, respectively. Close values follow from the graphs presented in Fig. 7a of the paper [6]: 0.3, 0.5, 0.7 and 1.0. These values do not differ significantly from our values for the PKN model. This indicates that the influence of the consistency and behavior indices of thinning fluids is quite similar for the KGD and PKN models.

From equation (29) it is obvious that to have the same fracture length at the reference time $t_r$, the starting speed of the fracture propagation should be less for a Newtonian fluid than that for a perfectly plastic fluid, while with growing time the speeds become firstly equal and then (near and after the reference time) the speed should become greater for the Newtonian fluid than for the perfectly plastic fluid. The quantitative description of the changes in the ratio $v_{*N}/v_{*p}$ is obtained by using the second of the equations (24):

$$\frac{v_{*N}}{v_{*p}} = \frac{\beta_{*N}}{\beta_{*p}} \frac{x_{*N}}{x_{*p}}, \quad (30)$$

or by taking into account for (29):

$$\frac{v_{*N}}{v_{*p}} = \frac{\beta_{*N}}{\beta_{*p}} \left(\frac{t}{t_r}\right)^{\beta_d}. \quad (31)$$

From (30) it is clear that the ratio of speeds differs from the ratio of lengths only by the multiplier $\beta_{*N}/\beta_{*p} = 5/6$. It is worth noting that equation (31) implies that the speeds become equal at the moment $t_v = t_r(\beta_{*p}/\beta_{*N})^{1/\beta_d} = 0.25476 t_r$. Obviously, the moment $t_v$ is actually the same for any thinning fluid under fixed reference values. This explains the fact, not mentioned in the papers [5, 6], that the graphs $v_*(t)$ for all thinning fluids intersect at the same point. It can be clearly seen in Fig. 8 of the paper [5] for the KGD model: the graphs $v_*(t)$ intersect at $t_v \approx 1600$ s. Our result for the same reference time ($t_r = 10^4$ s) is $t_v = 2548$ s, what is 1.6-fold greater. We see again reasonable agreement between the results for fluids with various consistency and behaviour indices despite the KGD and PKN models describe different plane-strain states.

We may also compare the fracture openings at the inlet $w(0,t)$ for various fluids. Since in the considered PKN model the net-pressure is proportional to the opening, the comparison refers to the net-pressure, as well. From the solution of the previous section it follows that

$$\frac{w_N(0,t)}{w_p(0,t)} = \frac{W_{0N}}{W_{0p}} \left(\frac{t}{t_r}\right)^{\beta_{wN}-\beta_{wp}}, \quad (32)$$

where $W_{0N} = 1.32628$, $W_{0p} = 1.44225$, so that $W_{0N}/W_{0p} = 0.91959$. The difference $\beta_{wN} - \beta_{wp}$ equals to the difference $\beta_d = \beta_{*N} - \beta_{*p}$, taken with the minus sign. Thus in view of (29), equation (32) may be written as

$$\frac{w_N(0,t)}{w_p(0,t)} = \frac{W_{0N}}{W_{0p}} \left(\frac{x_{*N}}{x_{*p}}\right)^{-1}. \quad (33)$$

We see that the ratio of openings is inverse proportional to that of lengths. Hence again the discussion of the ratio of lengths may be promptly interpreted in terms of the ratio of openings. Similar to the ratio of velocities, it follows from (32) that the openings become equal at the moment $t_w = t_r(W_{0N}/W_{0p})^{1/\beta_d} = 0.53328 t_r$. Again the moment $t_w$ is actually the same for any thinning fluid under fixed reference values. For $t_r = 10^4$ s, the time of intersection is $t_w = 5333$ s. We cannot compare this value with that corresponding to the KGD model, because the papers [5, 6] do not contain graphs for the dimensional opening.

The discussion shows that there are no decisive differences to choose between fluids with various behavior indices. At most, the differences may serve to have some quantity greater (less) at time notably less than the reference time. Therefore, the choice between fluids, which have various behavior indices while providing the same fracture length at the same reference time $t_r$, is to be made primarily from technological and/or economic considerations. Meanwhile, when using such considerations, one needs to know the consistency indices of the compared fluids, for which the fluids are equivalent in providing the same mechanical effect. The equation

(26) (or, equivalently, (27) or (28)) offers an answer. It gives the reference shear rate, which via equation (2) defines the consistency index of a fluid.

## 5. CONCLUSIONS

The conclusions of the paper are as follows.

(i) The analytical solution presented discloses general features of hydraulic fracturing with various *thinning* fluids. Specifically, for zero leak-off, the particle velocity is practically constant, while the modified opening is almost linear along the fracture. The self-similar fracture length $\xi_*$ is also practically independent on the fluid behavior index $n$ ($\xi_*$=0.040 for $n = 0$; $\xi_*$=1.001 for $n = 1$). This implies that for shear thinning fluids, the dependencies of the normalized quantities on the normalized time are universal. The differences occur mostly in the degrees of time $t^{\beta_*}$ and $t^{\beta_w}$, entering as multipliers to the self-similar fracture length and opening. The maximal difference in the exponent $\beta_*$ is 2/15 when comparing the limiting cases of a perfectly plastic ($n = 0$) and Newtonian ($n = 1$) fluids. The same maximal difference is in the exponent $\beta_w$.

(ii) Comparison of dimensional quantities for various fluids has sense only under an agreement on the expected (reference) shear strain rate (or, equivalently, shear strength). We suggest the criterion of equivalence of thinning fluids in their action in hydraulic fracturing: *fluids are assumed equivalent when at a prescribed reference time $t_r$ (say, treatment time) they produce fractures of the same length*. It is shown that the criterion provides the needed reference shear strain rate.

(iii) The differences in thinning fluids with various consistency and behavior indices appear mostly at time notably less than the reference time. Within the practically significant range of time, the ratio of the fracture lengths (propagation speed, opening, net pressure) at the same instances for various thinning fluids does no differ significantly. Under fixed reference time, the propagation speeds for all thinning fluids become equal at the same moment. Similar conclusion is true for the opening and the pressure at the inlet.

(iv) The influence of the consistency and behavior indices on the fracture length, particle velocity and propagation speed is quite similar for the PKN and KGD models despite the models refer to different plain-strain states. This suggests using the PKN model, which is simpler, for comparative studies of fluids with various viscous properties when accounting for leak-off and/or proppant movement.

(v) Differences between thinning fluids do not look significant from the mechanical point of view. They may be taken into account when wishing to have some quantity to be greater (less) at time notably less than the reference time. Consequently, a choice of a fracturing fluid is to be made mostly from technological and/or economic considerations under the condition that the compared fluids are equivalent in their mechanical effect. Equation (26) may serve for establishing the equivalence. It translates the equivalence in terms of the hydraulic fracture length and treatment time into the equivalence in terms of the properties of a fracturing fluid (behavior and consistency indices), expresses by equation (2).

## APPENDIX: ANALYTICAL SOLUTION IN SELF-SIMILAR VARIABLES

We seek the solution of the Cauchy problem (16), (17), (19), (20) in the Tailor's series:

$$Y(\xi) = C_Y \sum_{j=1}^{\infty} a_j \tau^j, \quad V(\xi) = V_* \sum_{j=0}^{\infty} b_j \tau^j, \tag{A1}$$

where $C_Y = \xi_*^{n+1} \beta_*^n / \alpha$. The BC (19) and the SE (20) give $a_1 = b_0 = 1$.

Substitution of the series (A1) into the dependence (17) yields

$$\sum_{k=0}^{\infty}(k+1)a_{k+1}\tau^k = \left(\sum_{j=0}^{\infty} b_j \tau^j\right)^n.$$

This provides recurrent evaluation of the coefficients $a_k$ ($k = 2, \ldots$) via the coefficients $b_j$ ($j = 1,\ldots, k - 1$). Specifically, for the first five coefficients we have:

$$a_1 = b_0 = 1, \quad a_2 = \frac{1}{2}nb_1, \quad a_3 = \frac{1}{6}n[(n-1)b_1^2 + 2b_2], \quad a_4 = \frac{1}{24}n[(n-1)(n-2)b_1^3 + 6(n-1)b_1b_2 + 6b_3],$$

$$a_5 = \frac{1}{120}n[(n-1)(n-2)(n-3)b_1^4 + 12(n-1)(n-2)b_1^2 b_2 + 24(n-1)b_1 b_3 + 24b_4]. \tag{A2}$$

The coefficients $a_k$ decrease faster than $1/k^2$ with growing $k$. Substitution of the series (A1) into the ODE (16) gives the second group of recurrent equations for $j \geq 2$:

$$b_j = -\frac{1}{j+\alpha}\left\{\sum_{k=2}^{j}(j-k+1+\alpha k)a_k b_{j-k+1} + (\alpha j - \frac{\beta_w}{\beta_*})a_j\right\}, \tag{A3}$$

with the starting values $a_1 = b_0 = 1$, $b_1 = \dfrac{1}{1+\alpha}\left(-\alpha + \dfrac{\beta_w}{\beta_*}\right)$.

Having the starting values $a_1$, $b_0$, $b_1$, we find $a_2$ from the second of (A2). Then (A3) provides $b_2$, the third of (A2) gives $a_3$, and so on. In the case of a Newtonian fluid ($n = 1$, $\alpha = 1/3$), we have $a_j = b_{j-1}/j$ ($j = 1,\ldots$) and for a constant influx ($\beta_q = 0$, $\beta_w = 1/5$, $\beta_* = 4/5$), the recurrence formulae (A3) reduce to those derived in [9]. For a perfectly plastic fluid ($n = 0$, $\alpha = 1/2$), all the coefficients $a_k$, $c_k$ are zero for $k > 1$. Then the solution for a constant influx is:

$$Y(\xi) = 2(\xi_* - \xi), \quad V(\xi) = V_*. \tag{A4}$$

The BC at the inlet (18) gives $\xi_* = (9/8)^{1/3}$. From (A4) we see that the function $Y(\xi)$ is linear in the self-similar distance from the inlet, while the self-similar particle velocity is constant along fracture. It is equal to the fracture speed $V_* = \xi_* \beta_*$, where $\beta_* = 2(1+\beta_q)/3$.

*Acknowledgement.* The author gratefully acknowledges the support of the FP7 Marie Curie IAPP transfer of knowledge program (project PIAP-GA-2009-251475).